\documentclass[conference]{IEEEtran}
\IEEEoverridecommandlockouts

\def\BibTeX{{\rm B\kern-.05em{\sc i\kern-.025em b}\kern-.08em
    T\kern-.1667em\lower.7ex\hbox{E}\kern-.125emX}}

\usepackage{amsmath,graphicx,multirow,url,times,booktabs,tabularx}
\usepackage{hyperref}
\hypersetup{
    colorlinks=true,
    linkcolor=blue,
    filecolor=magenta,
    urlcolor=cyan,
    pdftitle={Overleaf Example},
    pdfpagemode=FullScreen,
    }
\usepackage{cite}
\usepackage{color}
\usepackage{pdfpages}
\usepackage{amsfonts, amssymb}
\usepackage[caption=false]{subfig}
\usepackage{color}
\usepackage{float}
\usepackage{bbding}
\usepackage{siunitx}

\title{Disentangling Hierarchical Features for Anomalous Sound Detection Under Domain Shift}

\begin{document}

\author{
    \IEEEauthorblockN{
    Jian Guan$^{1}$\thanks{This work was partly supported by the Natural Science Foundation of Heilongjiang Province under Grant No. LH2022F010.}, 
    Jiantong Tian$^1$, 
    Qiaoxi Zhu$^2$, 
    Feiyang Xiao$^1$, 
    Hejing Zhang$^1$, 
    Xubo Liu$^3$
    }
\IEEEauthorblockA{
    $^1$Group of Intelligent Signal Processing, College of Computer Science and Technology, \\ Harbin Engineering University, Harbin, 150001, China\\
    $^2$Acoustics Lab, University of Technology Sydney, Ultimo, NSW 2007, Australia\\
    $^3$Centre for Vision Speech and Signal Processing, University of Surrey, Guildford, GU2 7XH, UK\\
    }
}

\maketitle
\begin{abstract}

Anomalous sound detection (ASD) encounters difficulties with domain shift, where the sounds of machines in target domains differ significantly from those in source domains due to varying operating conditions. Existing methods typically employ domain classifiers to enhance detection performance, but they often overlook the influence of domain-unrelated information. This oversight can hinder the model's ability to clearly distinguish between domains, thereby weakening its capacity to differentiate normal from abnormal sounds. In this paper, we propose a Gradient Reversal-based Hierarchical feature Disentanglement (GRHD) method to address the above challenge. GRHD uses gradient reversal to separate domain-related features from domain-unrelated ones, resulting in more robust feature representations. Additionally, the method employs a hierarchical structure to guide the learning of fine-grained, domain-specific features by leveraging available metadata, such as section IDs and machine sound attributes. Experimental results on the DCASE 2022 Challenge Task 2 dataset demonstrate that the proposed method significantly improves ASD performance under domain shift. 
\end{abstract}
\section{Introduction}
\label{sec:1}

Anomalous sound detection (ASD) identifies whether a machine's operating condition is normal or abnormal based on the emitted sound. It is typically an unsupervised learning task due to the difficulty in collecting diverse and rare anomalous sounds, leading to models being trained only on normal sounds \cite{DCASE2020, zhang2024first, ICASSP2023ContrastiveLearning, TWFR_GMM}. ASD becomes even more challenging in real-world applications due to the domain shift problem, where acoustic characteristics differ between the source domain and the target domain, leading to reduced performance \cite{DCASE2022}. 

Unlike traditional domain shift problems, which usually involve a single domain \cite{Luo_2019_CVPR_single_domain, Sankaranarayanan_2018_CVPR_single_domain}, ASD faces a multi-source domain shift problem \cite{Peng_2019_ICCV_multi_domain, Gan_2016_CVPR_multi_domain_attribute} caused by various factors in operating conditions, such as changes in machine operation modes, speeds, or environmental noise \cite{HMIC, DCASE2022, DCASE2021}. These variations create multiple domains within the source and target domains, complicating feature learning due to imbalanced sample sizes and diverse operating conditions, making effective ASD under domain shift particularly challenging.

\begin{figure*}[!t]
 \centering
 {
    \includegraphics[width=0.82\linewidth]{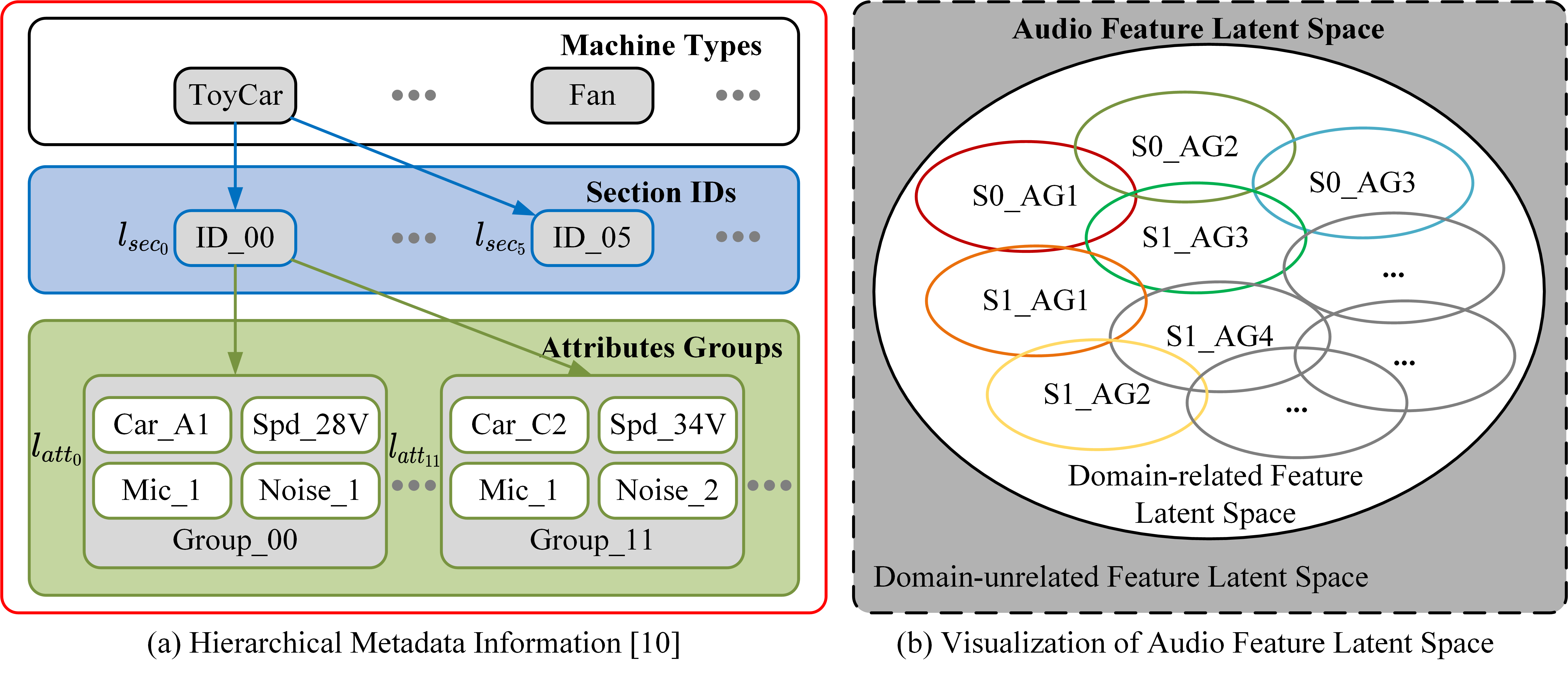}
 }
 \caption{Illustrations of (a) the hierarchical metadata structure accompanying the audio data for anomalous sound detection, and (b) the latent space of the audio data, containing both domain-related and domain-unrelated features. For example, S0\_AG1 corresponds to the latent space of audio labeled by section 00 and attribute group 1.}
 \label{fig:principle}
 \vspace{-2mm}
\end{figure*}

The DCASE 2022 Challenge Task 2 was designed to investigate domain shift in ASD \cite{DCASE2022}. The dataset for this task is organized hierarchically, encompassing machine types, section IDs, and attribute groups, as illustrated in Figure~\ref{fig:principle}(a). \textit{Section IDs} are defined by various \textit{attribute groups} that represent the machine's operational status and recording conditions. As shown in Figure~\ref{fig:principle}(a), a section ID comprises multiple attribute groups, such as ``Group\_00," which includes machine-related attributes (e.g., toy car speed ``Spd\_28V" and model ``Car\_A1") and recording-related attributes (e.g., microphone location ``Mic\_1" and environmental noise ``Noise\_1").
The changes in these attribute values primarily drive the domain shift amongst data of different section IDs.

\begin{figure*}[t]
    \centering
        \subfloat{
                 \includegraphics[width=0.9\linewidth]{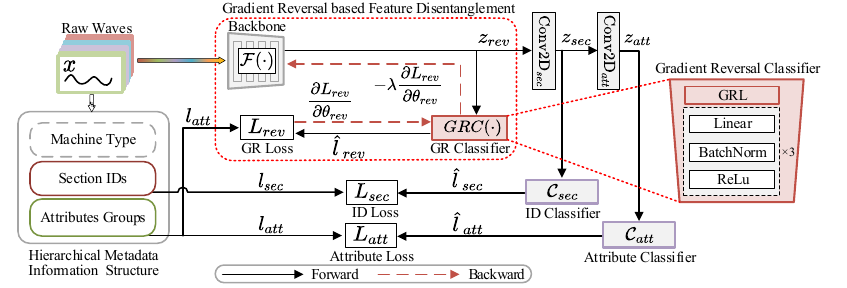}}
        \caption{The proposed GRHD method's training process. The gradient reversal classifier $GRC(\cdot)$ is used for feature disentanglement, and hierarchical metadata guides fine-grained, domain-specific audio feature learning. $\textbf{\textit{z}}_{rev}$ represents the disentangled domain-related feature, while $\textbf{\textit{z}}_{sec}$ and $\textbf{\textit{z}}_{att}$ correspond to section ID and attribute group features. $\mathcal{C}_{sec}$ and $\mathcal{C}_{att}$ are the classifiers for section ID and attribute group, producing labels $\hat{l}_{sec}$ and $\hat{l}_{att}$. The actual labels accompanying the audio data are $l_{sec}$ and $l_{att}$. The predicted label $\hat{l}_{rev}$ is from $GRC(\cdot)$.}
        \label{fig:model}
       \vspace{-5mm}
\end{figure*}

Although methods using section IDs and attributes as training labels for self-supervised classification can enhance detection performance, they often overlook the fact that attributes can affect machine sounds differently depending on the type of domain shift (i.e., section ID) \cite{DengTHU2022, WilkinghoffFKIE2022}. Instead, they tend to use section IDs and attributes in parallel for feature learning \cite{HMIC}. As a result, the interaction between attributes and domain shift types remains underexplored, despite its potential to improve ASD under domain shift.

In our previous study \cite{HMIC}, we introduced a Hierarchical Metadata Information Constraint (HMIC) structure to address the domain shift problem in ASD caused by changes in attribute values. This structure organizes metadata hierarchically, with section IDs representing types of domain shift and attribute groups (AGs) for specific domains. By grouping audio samples with the same attributes and values into AGs, we aimed to facilitate more effective learning of domain-related features. However, this approach may overlook the influence of domain-unrelated features as the grey area in Figure~\ref{fig:principle}(b), which can complicate the distinction between domain-related features and ultimately degrade ASD performance—a common issue in classification-based ASD methods.

This paper presents a Gradient Reversal-based Hierarchical Feature Disentanglement (GRHD) strategy for addressing domain shift in anomalous sound detection (ASD). The proposed method incorporates a gradient reversal layer within a hierarchical metadata constraint structure to hierarchically separate domain-unrelated features from domain-related features, thereby better disentangling overlapping latent features. Specifically, we introduce a gradient reversal classifier as a constraint to disentangle domain-unrelated and domain-related features during the coarse-grained audio feature learning across all attributes. Subsequently, a hierarchical metadata structure is employed as a further constraint to achieve more effective fine-grained domain-related audio feature learning from the disentangled coarse-grained domain-related features. In this process, section IDs and attribute groups are used as hierarchical constraints accounting for domain shift types and specific domains, respectively. Therefore, our proposed GRHD can achieve better coarse- and fine-grained audio feature representation for anomalous sound detection. Experiments on the DCASE 2022 Challenge Task 2 dataset \cite{DCASE2022} demonstrate that GRHD better disentangles domain-unrelated features and improves ASD performance under domain shift.

\section{Proposed Method}
\label{sec:2}

Figure~\ref{fig:model} illustrates the proposed method, which integrates a gradient reversal classifier into the hierarchical metadata constraint structure \cite{HMIC}. The gradient reversal classifier disentangles domain-unrelated from domain-related coarse features, while the hierarchical metadata constraint facilitates effective fine-grained domain-specific audio feature learning. The section ID and attribute group classifiers address domain shift types and specific domains, respectively.

\subsection{Gradient Reversal Based Feature Disentanglement}
\begin{table*}[!ht]
    \centering
    \caption{
    Performance comparison in terms of AUC (\%) and pAUC (\%) on the development set of the DCASE 2022 Challenge Task 2 dataset. 
    AUC-s, AUC-t and pAUC for Total performance are the harmonic averages of AUC in the source domains, AUC in the target domains, and pAUC across all machine types. HAUC is the harmonic average of AUC-s, AUC-t, and pAUC.
    }
    \label{table:performance_comparison}
    \resizebox{0.99\textwidth}{!}{
    \huge
    \begin{tabular}{@{}lccccccccccccccccccc@{}}
    \toprule
    \multirow {2}{*}{\textbf{Methods}} &
      \multicolumn{2}{c}{\textbf{ToyCar}} &
      \multicolumn{2}{c}{\textbf{ToyTrain}} &
      \multicolumn{2}{c}{\textbf{Bearing}} &
      \multicolumn{2}{c}{\textbf{Fan}} &
      \multicolumn{2}{c}{\textbf{Gearbox}} &
      \multicolumn{2}{c}{\textbf{Slider}} &
      \multicolumn{2}{c}{\textbf{Valve}} &
      \multicolumn{4}{c}{\textbf{Total}} \\
        \cmidrule(lr){2-3} \cmidrule(lr){4-5} \cmidrule(lr){6-7} \cmidrule(lr){8-9} \cmidrule(lr){10-11} \cmidrule(lr){12-13} \cmidrule(lr){14-15} \cmidrule(lr){16-19}
    &AUC &pAUC  &AUC &pAUC  &AUC &pAUC  &AUC &pAUC  &AUC &pAUC  &AUC &pAUC  &AUC &pAUC  &AUC-s &AUC-t &pAUC &HAUC  \\\midrule

    AutoEncoder \cite{DCASE2022} & \num{62.61} & \num{52.74} & \num{49.84} & \num{50.48} & \num{56.40} & \num{51.98} & \num{62.89} & \num{57.52} & \num{65.79} & \num{58.49} & \num{62.81} & \num{55.78} & \num{50.74} & \num{50.36} & \num{68.74} & \num{41.91} & \num{53.73} & \num{52.61} \\
    MobileNetV2 \cite{DCASE2022} & \num{55.54} & \num{52.27} & \num{51.58} & \num{51.52} & \num{60.26} & \num{57.14} & \num{59.49} & \num{56.90} & \num{62.70} & \num{56.03} & \num{51.69} & \num{54.67} & \num{62.16} & \num{62.42} & \num{63.80} & \num{50.04} & \num{55.65} & \num{55.94}\\
    JLESS \cite{BaiJLESS2022} & \num{82.50} & \num{64.68} & \num{64.52} & \num{55.60} & \num{79.83} & \num{61.04} & \num{67.28} & \num{67.25} & \num{86.77} & \num{72.67} & \num{86.73} & \num{71.41} & \(\mathbf{91.64}\) & \(\mathbf{80.01}\) & \num{82.85} & \num{73.96} & \num{66.69} & \num{73.92} \\
    Imp-Freq \cite{nguyen2024impact} & \num{65.60} & \num{56.10} & \num{55.70} & \num{52.50} & \num{63.10} & \num{56.90} & \num{74.30} & \num{61.70} & \num{75.50} & \num{64.60} & \num{86.10} & \num{74.50} & \num{71.50} & \num{62.90} & - & - & \num{61.31} & - \\
    DG-Mix \cite{nejjar2022dg_DGMIX} & \(\mathbf{87.51}\) & \(\mathbf{68.32}\) & \num{53.99} & \num{53.20} & \num{74.95} & \(\mathbf{64.75}\) & \(\mathbf{80.05}\) & \(\mathbf{70.74}\) & \num{85.85} & \num{71.22} & \num{80.94} & \num{68.91} & \num{83.51} & \num{71.65} & \num{77.55} & \(\mathbf{74.08}\) & \num{66.34} & \num{72.34} \\
    HMIC-AGC \cite{HMIC} & \num{74.20} & \num{57.99} & \num{62.24} & \num{54.02} & \num{73.45} & \num{59.61} & \num{58.21} & \num{55.83} & \num{76.94} & \num{67.66} & \num{83.52} & \num{69.11} & \num{71.32} & \num{66.93} & \num{77.46} & \num{61.68} & \num{61.06} & \num{65.93} \\
    \midrule
    \textbf{GRHD} & \num{85.20} & \num{64.13} & \(\mathbf{69.10}\) & \(\mathbf{58.91}\) & \(\mathbf{81.43}\) & \num{63.52} & \num{64.67} & \num{66.49} & \(\mathbf{89.45}\) & \(\mathbf{77.98}\) & \(\mathbf{88.99}\) & \(\mathbf{79.12}\) & \num{86.24} & \num{77.53} & \(\mathbf{84.64}\) & \num{72.43} & \(\mathbf{68.82}\) & \(\mathbf{74.72}\) \\ \bottomrule
    \end{tabular}}
\end{table*}
To disentangle domain-unrelated and domain-related features, a gradient reversal classifier, which includes a Gradient Reversal Layer (GRL) as shown in Figure~\ref{fig:model}, is introduced to constrain the model in learning coarse audio features across all attributes. Subsequently, attribute groups are utilized as self-supervised labels for the gradient reversal classifier.

Supposing $\textbf{\textit{x}} \in \mathbb{R} ^ {1 \times T}$ is the input audio signal with length of $T$, the disentangled feature embedding $\textbf{\textit{z}}_{rev}$ is extracted with
\begin{equation}
\label{eq:3}
    \textbf{\textit{z}}_{rev}= \mathcal{F}(\textbf{\textit{x}}),
\end{equation}
where $\mathcal{F}(\cdot)$ denotes the backbone \cite{HMIC}  designed based on spectral-temporal information fusion \cite{STgramMFN}.

During forward propagation, the gradient reversal classifier $GRC(\cdot)$ with gradient reversal predicts labels for input embeddings $\textbf{\textit{z}}_{rev}$, expressed as $\hat{l}_{rev} = GRC(\textbf{\textit{z}}_{rev})$. Due to imbalanced audio samples across different attribute groups, Focal loss \cite{lin2017focalloss} is utilized as the loss function $L_{Focal}$ to optimize the gradient classifier, with the ground truth label $l_{att}$ of the attribute group of the input audio, that
\begin{equation}
\label{eq:revLoss}
    L_{rev} = L_{Focal}(l_{att}, \hat{l}_{rev}).
\end{equation}

During backpropagation, the Gradient Reversal Layer (GRL) sends the opposite gradient back to the backbone, creating divergent learning objectives between the classifier and the backbone, indicative of adversarial learning. The parameters of the gradient reversal are defined as $\theta_{rev}$ for a simplified description\cite{gradient_reversal}, and the backpropagation process before and after the GRL is 
\begin{equation}
\label{eq:gradient reversal representation}
    \frac{\partial L_{rev}}{\partial \theta_{rev}} \gets GRL\left(\frac{\partial L_{rev}}{\partial \theta_{rev}}\right) = - \lambda \frac{\partial L_{rev}}{\partial \theta_{rev}},
\end{equation}
where $GRL(\cdot)$ denotes the function that refers to gradient reversal operations, and  $\lambda$ represents the intensity of gradient reversal, which becomes stronger as the training process.

By the adversarial learning between the classifier $GRC(\cdot)$ and the backbone, we constrain the backbone to extract unified non-domain-specific coarse features $\textbf{\textit{z}}_{rev}$ from the source and target domains to achieve feature disentanglement and alleviate the domain shift problem in ASD.

\subsection{Hierarchical Metadata Constrained Domain-Related Feature Learning}

For more fine-grained feature learning, following \cite{HMIC}, the hierarchical metadata structure is employed as a constraint to refine the coarse domain-related feature for more effective domain-specific feature learning. Here, the section ID and AG are employed as the constraint to learn the audio features $\textbf{\textit{z}}_{sec}$ and $\textbf{\textit{z}}_{att}$ related to domain shift type and specific domain, respectively. With the hierarchical metadata constrain, the audio features learnt from different levels are
\begin{equation}
\label{eq:5}
    \textbf{\textit{z}}_{sec} = \text{Conv2D}_{sec}( \textbf{\textit{z}}_{rev}),\ \ \ \textbf{\textit{z}}_{att} = \text{Conv2D}_{att}( \textbf{\textit{z}}_{sec}),
\end{equation}
where $\text{Conv2D}_{sec}$ and $\text{Conv2D}_{att}$ denote 2D convolutional layers used for audio feature extraction, following \cite{HMIC}. 

To employ section ID and AG as constraints for effective feature learning, two classifiers, ID classifier  $\mathcal{C}_{sec}(\cdot)$ and attribute classifier $\mathcal{C}_{att}(\cdot)$, are applied to obtain predicted labels, $\hat{l}_{sec} = \mathcal{C}_{sec}(\textbf{\textit{z}}_{sec})$ and $\hat{l}_{att} = \mathcal{C}_{att}(\textbf{\textit{z}}_{att})$, then the section ID label $l_{sec}$ and AG label  $l_{att}$ can be used to constrain the feature learning. Unlike the imbalanced issue in attribute classification, section classification does not present a similar problem. Therefore, deviating from \cite{HMIC}, we employ Focal loss instead of cross-entropy loss for attribute classification.
\begin{equation}
\label{eq:7}
    L_{sec} = CE(l_{sec}, \hat{l}_{sec}),~~~L_{att} = L_{Focal}(l_{att}, \hat{l}_{att}).
\end{equation}
Finally, the joint loss function $L_{total}$ for model learning is
\begin{equation}
\label{eq:9}
    L_{total} = \alpha L_{rev} + \beta L_{sec} + \gamma L_{att},
\end{equation}
where $\alpha$, $\beta$ and $\gamma$ are the penalty parameters. 

In summary, the proposed GRHD strategy incorporates a gradient reversal classifier within a hierarchical metadata constraint structure. This enhances domain-specific feature learning by increasing the distance between features of different domains in the latent feature space, thereby improving anomalous sound detection under domain shift.

\section{Experiment and Result}
\label{sec:3}

\subsection{Experimental Setup}
\label{sec:3_1}

\subsubsection{Dataset}

We conduct experiments on the development set of the DCASE 2022 Challenge Task 2 dataset\footnote{\url{https://dcase.community/challenge2022}} to evaluate our method. Domain shift problems are defined in each section with different attribute values. In addition, there are \num{990} and \num{10} audio samples in the source and target domains, respectively. So, we can evaluate the domain generalisation performance of the ASD method on this dataset.

\subsubsection{Implementation}

We set the frame size as \num{1024} with an over-lapping of \num{50}\% to obtain the Log-Mel spectrograms, and the number of Mel filter banks as \num{128}. We employed the Adam Optimizer \cite{Adam} with a learning rate of \num{0.001} for model training and the cosine annealing strategy for learning rate decay. The model is trained with \num{150} epochs for each machine type. We empirically selected the penalty parameters ($\alpha$, $\beta$, and $\gamma$ in Eq.~\ref{eq:9}) for each machine type by searching the parameters resulting in the best performance.

\subsubsection{Performance Metrics}

The evaluation metrics are the area under the receiver operating characteristic curve (AUC) and the partial-AUC (pAUC), expressed as percentages, with higher values indicating better performance, following the DCASE 2022 Challenge Task 2 \cite{DCASE2022}.

\subsection{Experimental Results}
\label{sec:3_2}

\subsubsection{Performance Comparison}
Table~\ref{table:performance_comparison} shows the comparison of our proposed GRHD method with the other methods for ASD under domain shift, where AutoEncoder \cite{DCASE2022} and MobileNetV2 \cite{DCASE2022} are the official baseline methods for DCASE 2022 Challenge Task 2. JLESS \cite{BaiJLESS2022}, Imp-Freq \cite{nguyen2024impact} and DG-Mix \cite{nejjar2022dg_DGMIX} adopt Mixup strategy \cite{zhang2017mixup} for generalized feature learning to address the domain shift problem. HMIC-AGC \cite{HMIC} is the backbone of the proposed GRHD method. 

We can see that GRHD significantly outperforms the DCASE baseline systems in all evaluation metrics, obtains the performance improvement compared with the HMIC-AGC backbone, and achieves the best overall performance among all systems in HAUC for ASD under domain shift. By extracting the domain-related coarse and fine features, our method is better than other methods with the Mixup strategy for generalized audio feature learning to address anamlous sound detection under domain shift.

\begin{table}[!htbp]
    \vspace{-2mm}
    \centering
    \caption{Ablation study of the proposed GRHD method for ASD under domain shift. $GRC(\cdot)$, $\mathcal{C}_{sec}$ and $\mathcal{C}_{att}$ refer to the gradient reversal classifier, section ID classifier, and attribute classifier, respectively.}
    \vspace{-2mm}
    \label{table:ablation_classifier}
    \resizebox{0.48\textwidth}{!}{
    \begin{tabular}{@{}ccccccc@{}}
    \toprule
    \hspace{0.8mm} $GRC(\cdot)$ \hspace{0.8mm} & \hspace{0.8mm} $\mathcal{C}_{sec}$ \hspace{0.8mm} & \hspace{0.8mm} $\mathcal{C}_{att}$ \hspace{0.8mm} & \hspace{0.8mm} \textbf{AUC-s} \hspace{0.8mm} & \hspace{0.8mm} \textbf{AUC-t} \hspace{0.8mm} & \hspace{0.8mm} \textbf{pAUC} \hspace{0.8mm} & \hspace{0.8mm} \textbf{HAUC} \hspace{0.8mm} \\ \midrule
    \XSolidBrush & \Checkmark & \XSolidBrush & \num{73.60} & \num{56.05} & \num{57.51} & \num{61.45} \\
    \XSolidBrush & \XSolidBrush & \Checkmark & \num{77.77} & \num{60.99} & \num{60.50} & \num{65.53} \\
    \XSolidBrush & \Checkmark & \Checkmark & \num{77.46} & \num{61.68} & \num{61.06} & \num{65.93} \\
    \Checkmark & \Checkmark & \Checkmark  & \(\mathbf{84.64}\) & \(\mathbf{72.43}\) & \(\mathbf{68.82}\) & \(\mathbf{74.72}\) \\ 
    \bottomrule
    \end{tabular}}
    \vspace{-2mm}
\end{table}

\subsubsection{Ablation Study}
To validate the impact of different classifiers on ASD performance, we conduct an ablation study to verify these three classifiers $GRC(\cdot)$, $\mathcal{C}_{sec}$ and $\mathcal{C}_{att}$ in our model.
The results are given in Table~\ref{table:ablation_classifier}, where we can see the proposed gradient reversal strategy contributes the most to ASD under domain shift. 

Specifically, it provides nearly 11\% improvement in the target domain compared to our method without gradient reversal.
This indicates that our proposed method can effectively distinguish the boundaries between domain-related and domain-unrelated features by disentangling attribute related features from audio features.
The ablation experiments also showed that using attribute information can more effectively perform ASD than using section information. This verifies
the change of attribute values is the reason for domain shift, and using attribute values as self-supervised labels can extract domain-related audio features, thereby alleviating the domain shift problem in ASD.

\subsubsection{Visualization Analysis}
\begin{figure}[!htbp]
 \centering
 \subfloat[\centering{Without gradient reversal strategy}]{
 \includegraphics[width=0.85\columnwidth]{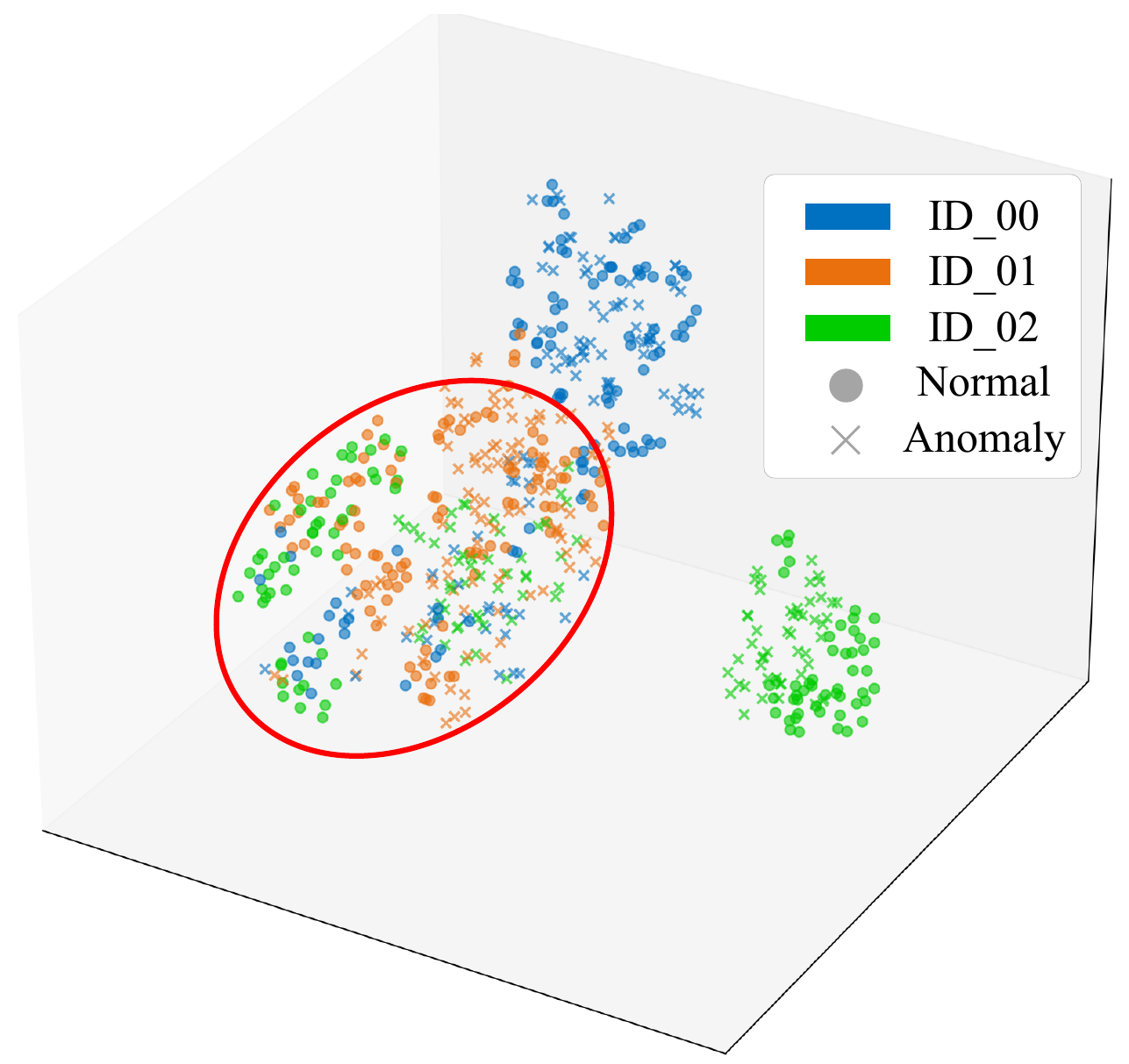}
     }
 \quad
 \subfloat[\centering{With gradient reversal strategy}]{
 \includegraphics[width=0.85\columnwidth]{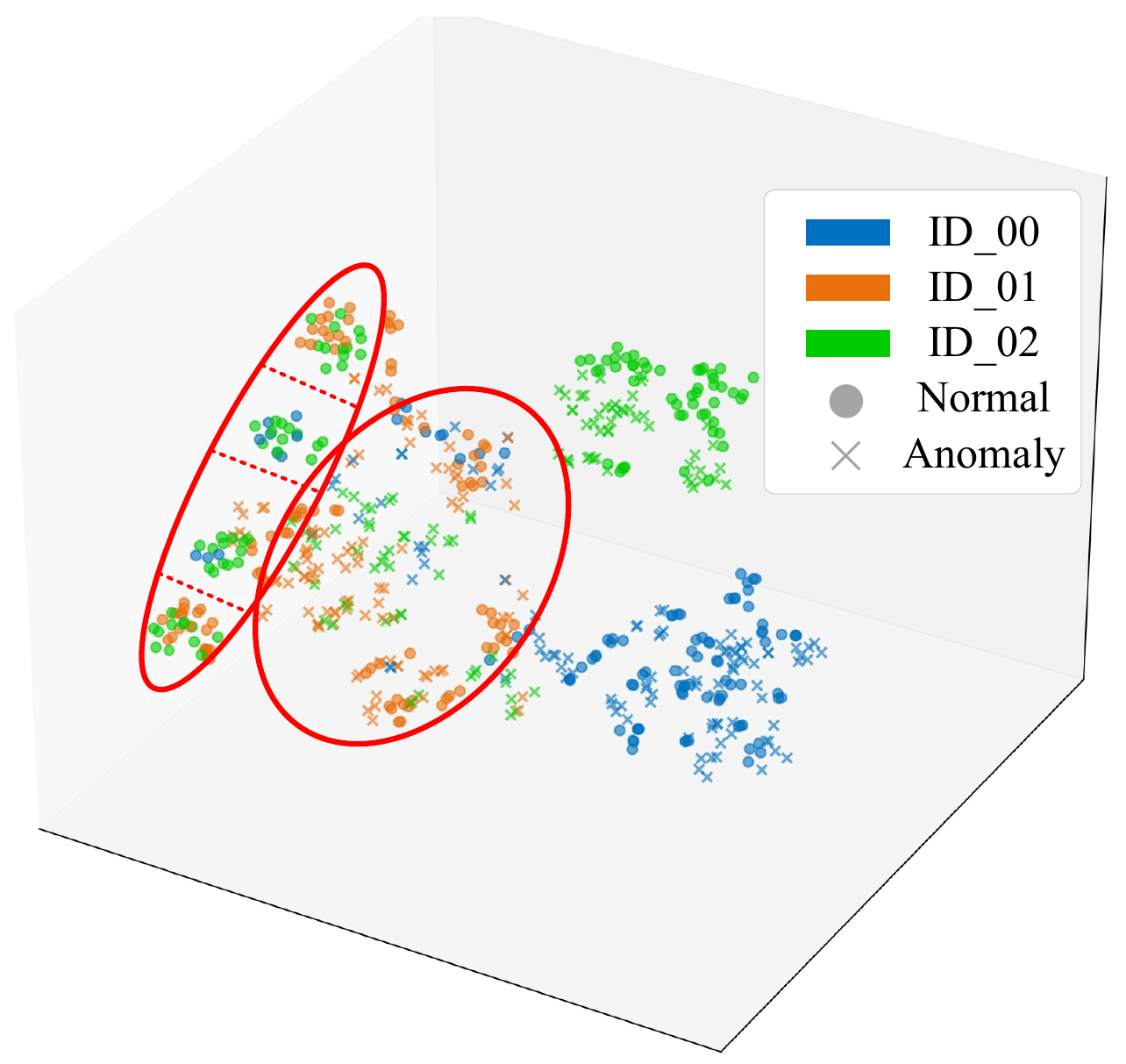}
     }
 \caption{The t-SNE visualization of the latent features with or without gradient reversal strategy for machine type ToyCar. Different colours represent different section IDs. “$\bullet$” and “$\times$” represent normal and anomalous sounds, respectively.}
 \label{fig:tsne}
 \vspace{-4mm}
\end{figure}

To further verify the effectiveness of our proposed method, we provide 3D t-SNE \cite{tSNE} cluster visualization of our method without and with gradient reversal strategy, as shown in Figure~\ref{fig:tsne} (a) and (b) respectively. With the gradient reversal strategy, latent features across different sections are better aggregated, as highlighted by the red elliptical box with dashed lines in Figure~\ref{fig:tsne}(b), rather than being completely mixed. Moreover, the overlap between normal and anomalous features is significantly reduced, as illustrated by the other red elliptical boxes in Figure~\ref{fig:tsne}, indicating that the gradient reversal strategy can effectively distinguish domain-related features from domain-unrelated features.

\section{Conclusion}
\label{sec:4}

In this paper, we have presented a gradient reversal-based hierarchical feature disentanglement strategy for anomalous sound detection under domain shift. The proposed gradient reversal strategy effectively disentangles domain-related features from complex audio features. Moreover, by incorporating a hierarchical metadata structure as the constraint to consider the types of domain shift and specific domain, we can obtain more effective fine-grained features, significantly enhancing anomalous sound detection performance under domain shift, as confirmed by experiments on the DCASE 2022 Challenge Task 2 dataset.

\bibliographystyle{IEEEtran}
\bibliography{refs}

\end{document}